# Solitary vortices supported by localized parametric gain

Changming Huang,[1] Fangwei Ye,[1,*] Boris A. Malomed,[2] Yaroslav V. Kartashov,[3] and Xianfeng Chen[1]

[1]State Key Laboratory of Advanced Optical Communication Systems and Networks, Physics Department, Shanghai Jiao Tong University, Shanghai 200240, China
[2]Department of Physical Electronics, School of Electrical Engineering, Faculty of Engineering, Tel Aviv University, Tel Aviv 69978, Israel
[3]Institute of Spectroscopy, Russian Academy of Sciences, Troitsk, Moscow Region, 142190, Russia
*Corresponding author: fangweiye@sjtu.edu.cn

We demonstrate the existence and stability of bright vortex solitons sustained by a ring-shaped parametric gain, for both focusing and defocusing Kerr nonlinearities in lossy optical media. With the defocusing nonlinearity, the vortices are stable at all values of the detuning parameter, while under the focusing nonlinearity their stability region is limited to some positive values of the detuning. Unstable vortices in the focusing medium transform into stable rotating azimuthons.

Light beams carrying phase singularities, i.e., vortices, appear in many branches of optics. Localized vortices form when the material nonlinearity is in balance with diffraction [1-3]. However, in conservative uniform nonlinear media vortex solitons are usually prone to azimuthal instabilities [4-6]. The instabilities can be eliminated in nonlocal nonlinear media [4,7-11], by competing nonlinearities [12,13], or by means of spatially inhomogeneous distributions of the refractive index [14-16] and nonlinear coefficients [17-20].

Vortex solitons have also been studied in various non-conservative systems [21-25], when the additional balance between the optical gain and losses takes place. In most cases, dissipative solitons, including vortices, were constructed with the help of the linear and nonlinear gain which is not sensitive to the phase of the light beam. Effects of the linear *parametric gain*, which couples to the phase of the input signal, were also studied in various spatially uniform settings [26-32]. It produces amplification of the signal due interaction of its conjugate form with the external field which provides the power signal. Very recently, we introduced a one-dimensional dissipative nonlinear system with *spatially localized* parametric gain, and demonstrated that it provides for a specific mechanism of the stabilization of solitons [33].

In this Letter, we introduce spatially localized parametric gain in the two-dimensional geometry, and demonstrate that it can support vortex solitons, provided that the phase of gain is matched to expected vortical structure. In the focusing medium, all the vortex solitons are unstable if the light propagation is exactly tuned to the parametric gain, while a finite stable region appears for positive detuning. In the case of the defocusing nonlinearity, solitary vortices can be stable for all values of the detuning. We also demonstrate that unstable vortices propagating in the focusing medium split into several fragments, transforming themselves into rotating azimuthons.

Thus, we consider the propagation of the two-dimensional optical beam along the $\xi$ axis in Kerr medium with the parametric gain localized in a ring-shaped area. The evolution of the normalized field amplitude $q$ is governed by the extended nonlinear Schrödinger equation:

$$i\frac{\partial q}{\partial \xi} = -\frac{1}{2}\left(\frac{\partial^2 q}{\partial \eta^2} + \frac{\partial^2 q}{\partial \zeta^2}\right) - (k+i\gamma)q + \sigma|q|^2 q - ar^{2m}\exp(2im\varphi - r^2/d^2)q^* \quad (1)$$

Here the transverse coordinates $\eta, \zeta$ and propagation distance $\xi$ are normalized to the characteristic transverse scale $r_0$ and diffraction length $L_{\text{diff}} = 2\pi n r_0^2/\lambda$, respectively. $r = (\eta^2 + \zeta^2)^{1/2}$ is the radial coordinate, $\varphi$ is the azimuthal angle, $\gamma > 0$ is the strength of the linear losses, and $\sigma = -1(+1)$ corresponds to the focusing (defocusing) Kerr nonlinearity. The last term of Eq. (1) defines the ring-shaped parametric gain, which is assumed to be introduced by a pump beam carrying topological charge $2m$. Parameters $a$, $d$, and $k$ define the amplitude, spatial width, and detuning of the parametric gain (the latter parameter is proportional to the wavenumber mismatch $2k_0(\omega) - k_0(2\omega)$ arising in the description of standard parametric interaction of waves with carrying frequencies $\omega$ and $2\omega$), the asterisk standing for the complex conjugate. Thus, here we assume that gain is provided by the parametric interaction between fundamental frequency and second-harmonic waves, use a non-depleted approximation for the second-harmonic wave, whose amplitude determines the gain parameter $a$, and assume that fundamental-frequency wave experiences strong

$\chi^{(3)}$ nonlinearity of the material. For the beam with characteristic width $r_0 = 40\ \mu m$ at wavelength $\lambda = 1.55\ \mu m$, propagating in a material with the refractive index $n \simeq 1.5$ and nonlinearity coefficient $n_2 = 3 \times 10^{-14}\ cm^2/W$, $q \sim 1$ corresponds to the field intensity $\sim 1\ GW/cm^2$, and $\gamma \sim 1$ implies the absorption coefficient $\sim 1\ cm^{-1}$.

We look for stationary vortex-soliton solutions to Eq. (1) by assuming $q(r,\varphi,\xi) = w(r)\exp(im\varphi)$, where $\eta = r\cos\varphi$, $\zeta = r\sin\varphi$. This ansatz carries vorticity $m$ which is matched to the topological charge of the parametric gain, as otherwise no vortex solutions can be constructed. The substitution of the ansatz into Eq. (1) leads to the following equation for the radial profile:

$$-\frac{1}{2}\left(\frac{\partial^2 w}{\partial \eta^2}+\frac{\partial^2 w}{\partial \zeta^2}\right)-(k+i\gamma)w+\sigma|w|^2 w - ar^{2m}\exp(-r^2/d^2)w^* = 0, \quad (2)$$

which can be solved by means of the Newton's iteration method. The scales are fixed by setting $\gamma = 1$. We also fix the width of gain landscape by setting $d = 1$, while the mismatch and strength of the parametric drive, $k$ and $a$, will be varied.

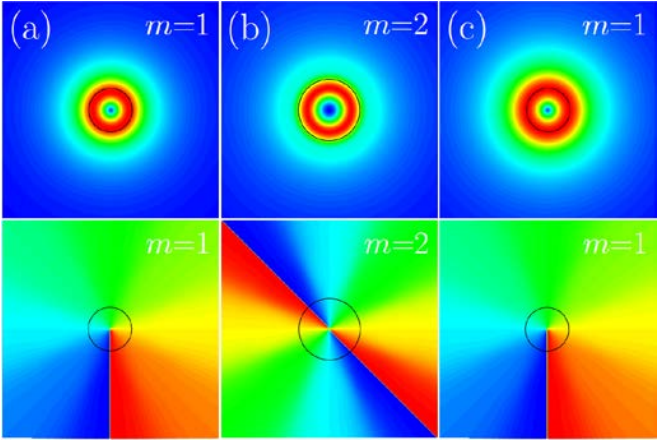

Fig. 1. (Color online) Distributions of the absolute value of the field (top) and phase (bottom) for vortex solitons with (a) $m=1$, $k=1.3$, $a=4.9$, (b) $m=2$, $k=3$, $a=4.7$, and (c) $m=1$, $k=3$, $a=7$. Panels (a) and (b) correspond to the focusing medium, while (c) represents the defocusing one. Black circles indicate the radius at which the parametric gain has its maximum.

Figure 1 shows typical profiles of single-charged ($m=1$) and double-charged ($m=2$) vortex solitons. The parametric gain can support solitary vortices in the focusing and defocusing media alike. In the plots it is seen that the radius of the vortex soliton approximately, but not exactly, coincides with the radius of the gain-carrying ring. The properties of these vortices are summarized in Fig. 2, where the dependence of energy flow $U = \int_{-\infty}^{\infty}\int_{-\infty}^{\infty} w^2 d\eta d\zeta$ on gain amplitude $a$ is plotted at two different values of the detuning $k$. At each value of $k$, curves are plotted either for $m=1$ or for $m=2$, but not for both, as the curves for $m=1$ and $m=2$ with the same $k$ are very similar to each other, except for a horizontal shift. It is worthy to note that, at $k=3$ [Fig. 2(a)], the curve for the focusing system monotonically increases, while its counterpart for the defocusing system initially goes to the left and then turns to the right at a threshold value of the gain, $a = a_{th}$. In contrast to that, at relatively small $k=1.8$ [Fig. 2(b)], the curve for the focusing system initially goes to the left and then turns around, while its counterpart in the defocusing medium monotonically increases. In the limit of $U \to 0$, Figs. 2(a) and 2(b) also demonstrate that the solitons in both the focusing and defocusing media appear at the same initial value of the gain, $a = a_g$, which implies the existence of a localized "gain-guided" vortex mode in the corresponding linear system.

Figures 2(c) and 2(d) present the dependence of $a_{th}$ on detuning $k$, for both focusing and defocusing nonlinearities. The figures show that there exists a point, $k = k_g$ [$k_g \approx 1.82$ in Fig. 2(c) and $k_g \approx 2.48$ in Fig. 2(d)], where the curves for the focusing and defocusing nonlinearities cross each other. Numerical results demonstrate that, at this point, both respective curves $U(a)$ go directly to the right, which means that in this point $a_{th} = a_g$ and lower branch of solutions does not exist. Actually, at this specific point, the numerical results demonstrate that the $U(a)$ curves for the focusing and defocusing media are indistinguishable at low powers, and the corresponding vortices are very similar in their widths, amplitudes, and profiles. As soon as $k$ shifts from $k_g$ value, the $U(a)$ curve for either focusing or defocusing media (but not both curves) develops a small segment with the negative slope (lower branch of solutions). Thus, two $U(a)$ curves split as soon as one moves away from the linear limit. Figures 2(c) and 2(d) show too that the threshold value $a_{th}$ for the vortex formation in the focusing medium is smaller than in the defocusing one at $k < k_g$, and the situation is reversed at $k > k_g$.

Next, we address the stability of the vortices. We have performed systematic simulations of Eq. (1) with the input corresponding to perturbed vortices, i.e., $q|_{\xi=0} = w(\eta,\zeta)\exp(im\varphi)[1+\rho(\eta,\zeta)]$, where $\rho$ is a random or a regular small perturbation, $|\rho(\eta,\zeta)| \ll |w(\eta,\zeta)|$. Simulations have revealed that, all the vortices in the defocusing medium are stable if $U(a)$ is a monotonously increasing function (i.e., if $k \geq k_g$), see Fig. 2(b), or they are stable at the branch with the positive slope if $U(a)$ has an initial small negative-slope segment, see Fig.2 (a). Therefore, the vortices in the defocusing medium with topological charges $m=1,2$ can be stable at all values of detuning $k$.

In sharp contrast with the defocusing case, the vortex stability under the focusing nonlinearity strongly depends on the detuning $k$. First, all the vortices are *unstable* at $k \leq 0$. Stable vortices appear only in limited intervals of gain strength at $k > 0$, as shown in Fig. 2(c) and Fig. 2(d), where the stability domain is the one defined by $a_{th} \leq a \leq a_{cr}$, where $a_{cr}$ is a critical value of gain strength $a$ above which the vortex solitons are no longer stable. If we define the width of stability domain as $\delta a = a_{cr} - a_{th}$, one clearly sees from the plots that the stability domain initially broadens, reaches its maximum, then shrinks, and finally disappears. Notice that the stability regime covers the $k = k_g$ point. For $k < k_g$, the stability regime corresponds to the initial segment of the monotonic $U(a)$ curve [Fig.2(a)], whereas for $k > k_g$ the stability regime corresponds to the initial segment of the upper branch of the nonmonotonic $U(a)$ curve in Fig. 2(b). It is worthy to note that even the doubly-charged vortices have a wide stability region.

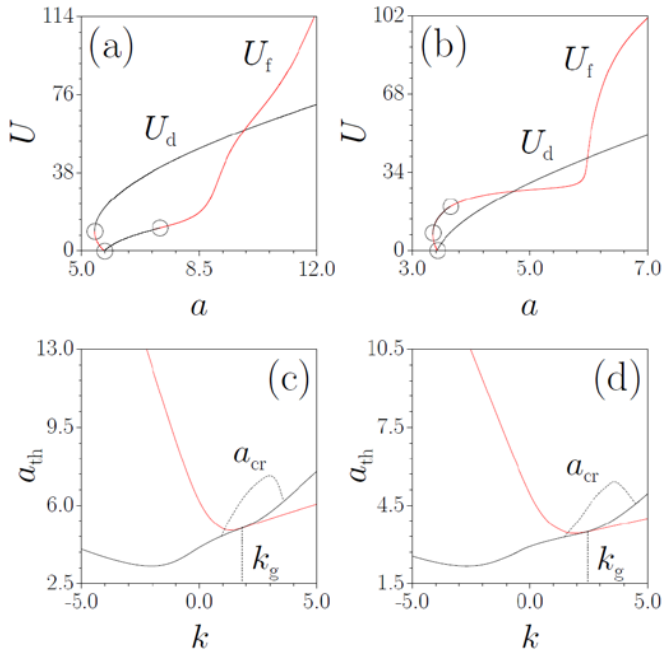

Fig. 2. (Color online) The energy flow for $m = 1$ (a) and $m = 2$ (b) vortex solitons in the focusing (curves labeled $U_f$) and defocusing (curves labeled $U_d$) media versus the gain strength, $a$. Panel (a) and (b) correspond to $k = 3$ and $k = 1.8$, respectively. The left, central, and right circles indicate, respectively, the threshold gain $a_{th}$ (left), the value $a_g$ corresponding to gain-guiding in the linear medium (center), and critical gain $a_{cr}$ (right). The black and red segments represent stable and unstable solution branches. The threshold and critical gain are shown versus $k$ for $m = 1$ in (c), and for $m = 2$ in (d). The black and red lines show $a_{th}$ in the focusing and defocusing media, respectively. The solitary vortices in the focusing medium are stable at $a_{th} < a < a_{cr}$.

Figure 3 displays typical propagation dynamics of vortex solitons in the focusing medium. The unstable solitons belonging to the branch where $dU/da > 0$ [see Figs. 2(a) and 2(b)] usually break into two (for $m = 1$, the first row) or three (for $m = 2$, the second row) fragments as a result of development of modulation instability in the azimuthal direction, which is a standard instability mechanism leading to decay of ring-like excitations in the focusing nonlinear medium. However, due to radial confinement of parametric gain the fragments do not separate in the course of the subsequent propagation. Instead, they form apparently stable complexes rotating at a constant angular velocity. Such rotating structures may be considered as a kind of azimuthons [16] supported by the localized parametric gain. The solitons belonging to the unstable branch where $dU/da < 0$ simply decay in the course of propagation without considerable broadening, suggesting an exponential decay instability. Stable vortices quickly shed off perturbations and then propagate without any change in the shape over an indefinitely long distance (the third row).

The instability scenario for vortex solitons is quite different in the defocusing medium: they quickly decay and eventually disappear (not shown here in detail). However, the unstable vortex does not notably diffract or shrink in the course of the decay.

Summarizing, we have studied the existence, shapes and stability of vortex solitons supported by the ring-shaped parametric gain, which carries the phase structure that makes it possible to sustain a parametrically driven vortex. We have found that the vortices have an infinitely wide stability region in the case of the defocusing nonlinearity, so that they may be stable at all values of the detuning. On the contrary, under the focusing nonlinearity the vortices are stable in a relatively narrow domain at positive values of the detuning. The unstable vortices in the system with the focusing sign of the nonlinearity transform themselves into robustly rotating azimuthons.

The work of C. Huang and F. Ye is supported by the NNSFC (11104181).

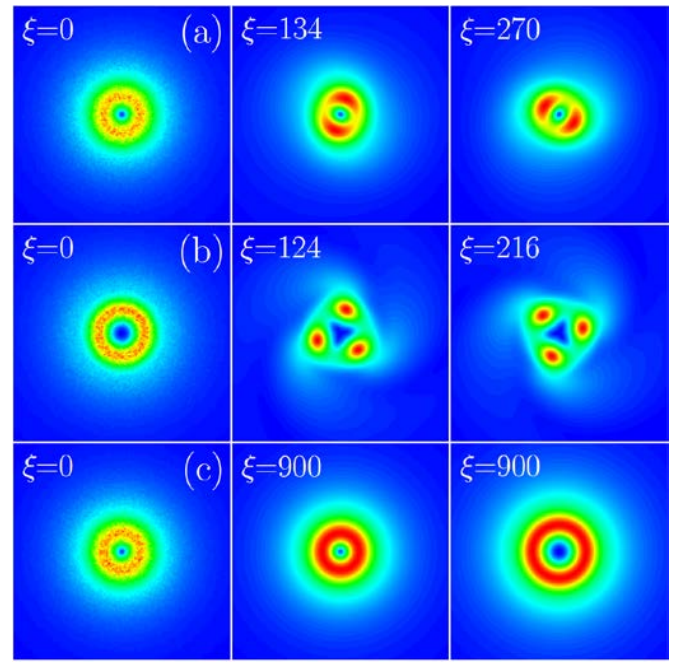

Fig. 3. (Color online) The evolution of profiles of the absolute value of the field in the course of the propagation in the focusing medium. (a) The breakup of a perturbed unstable vortex with $m = 1$ at $k = 1.3$, $a = 5.5$; (b) the breakup of an unstable vortex with $m = 2$ at $k = 1.8$, $a = 4.5$; (c) stable propagation of perturbed vortex with $m = 1$ at

$k=1.3$, $a=4.9$ (left and center), and of the one with $m=2$ at $k=1.8$, $a=3.6$ (right).